# Collective Coordinates Method in Relativistic Field Theory

A.V. Shurgaia
Department of theoretical physics
A. Razmadze Mathematical Institute
at Ivane. Javakhishvili State University
2 Merab Aleksidze 2nd Lane
0193 Tbilisi Georgia

**Abstract**
A field theoretical perturbation theory in inverse powers of coupling constant is developed which is manifestly covariant in every order of the expansion. A dilatation operator serves as an evolution dynamical one in a scale non-invariant theory.

## 1. Introduction

The quantization of interacting fields with nontrivial solutions of the classical equations of motion requires a knowledge of the correct wave function of the ground state of the overall system. Symmetry principles can be a great help to solve this problem. These principles constitute the basis of the strong coupling method (or the method of collective coordinates, as it is now called), which was first proposed by G. WENTZEL, W. PAULI, and S. DANCOFF [l] in an investigation of the fixed source model in the strong coupling limit. A mathematically rigorous and consistent formulation of this method was given by N. N. BOGOLUBOV [2] in 1950. More recently this method attracted attention in studies of nonlinear classical field theories, and in the papers of many authors, it was developed for interacting systems with arbitrary symmetries and applied to various quantum field theoretical models with either the canonical method or the path integral formulation of the field theory [3-7]. An alternative procedure was proposed in ref. [S] in which the collective variables are introduced in the quantum Yang-Feldman equations. One should note, that the authors of these papers have used nonrelativistic symmetry group parameters as the collective coordinates. In principle, the corresponding consideration of relativistic symmetry parameters is impossible in ordinary quantum field theory. Indeed, it is well known, that from a relativistic point of view, all physical quantities can be subdivided into dynamical i. e. interaction-dependent quantities and kinematical quantities [9]. The strong coupling method makes it possible to construct a perturbation theory for the dynamical quantities in such a way, that their eigenvalue spectrum can be described by an exact representation of the symmetry group. In standard field theory, constructed on the $t = const.$ surface, only the boost and energy are dynamical and we can describe them by the invariants of the three-dimensional Euclidean motion group (and an internal symmetry group, if present). Thus it is impossible to introduce Lorentz group parameters as collective variables into ordinary field theory in the sense mentioned above. To solve this problem we have to drop the definition of absolute time and construct the field theory on a surface, on which the Lorentz transformations carry the kinematical information. The generalized Hamiltonian formalism, developed by DIRAC [l0], opens, we think, the possibility to construct such a theory. The connection between the strong coupling method and the generalized Hamiltonian approach was first pointed out in ref. [ll]. In ref. [12] the general theory of the strong coupling method is developed for a nonrelativistic particle, interacting with the quantum field and for an arbitrary symmetry. A similar procedure is adopted in ref. [13] in a study of relativistic symmetry with the help of the strong coupling method. With this aim in mind, the author has constructed the theory on an arbitrary space-like surface, choosing the gauge at the flat surface, thereby losing the exact covariance. The relativistic symmetry has been investigated in ref. [14], in particular, the Yang-Feldman equation is studied in the two-dimensional scalar theory. The results the authors obtained correspond to the quantization on the flat surface. The covariant generalization of the strong coupling method with the application has been given in ref. [l5]. The strong coupling transformations are defined in such a way, that the field operators become at the same time operators in the space of the Poincare group generator algebra. These generators are then equated to the corresponding physical quantities constructed with the Noether theorem. Then Dirac-type constraints arise. In ref. [16] an alternative procedure is proposed that treats Lorentz invariance as an exact symmetry. In particular, the field theory is constructed on the surface, on which the Lorentz transformations carry the kinematical information, and the perturbation theory is developed, with exact covariance in each order. In what follows, we present a revised version of this approach for arbitrary strongly interacting fields.

## 2. Generalized Hamiltonian Dynamics

In this section, we recall some basic concepts of generalized Hamiltonian dynamics on an arbitrary space-like surface [l0]. Any three-dimensional surface is defined by specifying the four coordinates at any point on it as functions of three parameters $u_r$: $x_\mu = x_\mu(u_r), r = 1, 2, 3$. The space-likeness of the surface implies the existence of the normal time-like vector $\eta_\mu(u_r)$ at every point on it: $\eta_\mu \eta^\mu = 1, \eta_\mu \frac{\partial}{\partial u^r} \eta_\mu = 0$. The surface metric is given by the relation:
$$ds^2 = dx_\mu dx^\mu = \gamma^{rs} u_r u_s,$$
with
$$\gamma^{rs} = \frac{\partial x_\mu}{\partial u_r} \frac{\partial x^\mu}{u_s}.$$
is the metric tensor on it. Since $ds^2 < 0$, $\det \gamma^{rs} \equiv -\Gamma^2 < 0$. An inverse tensor is defined by the equality $\gamma^{rs} \gamma_{\sigma t} = \delta^r{}_t$. Any four-vector $a_\mu$ can be decomposed with respect to the coordinate system into normal and contravariant parts: $a_\perp = a_\mu \eta^\mu$ and $a^r = a_\mu \frac{\partial \eta_\mu}{\partial u_r}$. respectively. A covariant component is defined by the relation $a_r = \gamma_{rs} a^s$.

We can consider the quantities $x_\mu(u_r)$, defined in this way, as dynamical variables, which can also be unphysical. We next introduce a parameter $\tau$ (which is monotonically increasing with t), t), which varies from one surface to the other.
$$V_\mu = \pi_\mu(\tau, u_r) + K_\mu \approx 0,$$
where $\pi_\mu(\tau, u_r)$ are the canonical momenta conjugate to $x_\mu(\tau, u_r)$ and the functions $K_\mu$ do not contain them. The Lagrangian becomes a homogeneous function of the velocities, and the canonical Hamiltonian vanishes. The Hamiltonian now is a linear combination of the constraints, i.e.
$$H' = \int d^3 u c_\mu(\tau, u_r) V^\mu.$$

The equations of motion for an arbitrary function of the canonical variables take the form:

$$\dot{f} = \frac{\partial f}{\partial \tau} + \int \delta^3 u c_\mu(\tau, u_r) \{f, V^\mu\},$$

where
$$\dot{x}_\mu(\tau, u_r) = c_\mu(\tau, u_r).$$

Thus, the quantities $c_\mu(\tau, u_r)$ turn out to be the velocities $\dot{x}_\mu(\tau, u_r)$ and, consequently, a choice of $c_\mu(\tau, u_r)$ defines the variation of the surface variation depending on $\tau$. Suppose now the $c_\mu(\tau, u_r)$ are $\tau$-independent functions. Then we get the following representation for the $x_\mu(\tau, u_r)$:

$$x_\mu(\tau, u_r) = x_{0\mu}(\tau, u_r) + c_\mu(u_r)\tau.$$

The equations of motion for z,(u, z) then assume the form:

$$\dot{\pi}_\mu(\tau, u_r) = -c_\nu(u_r)\frac{\partial K^\nu}{\partial x^m}.$$

Via $K^\mu$, these equations involve the interaction terms and, consequently, depend on them in some complicated manner. Turning now to the field variables we can conclude, that the invariance of the Euler-Lagrange equations under space-time translations makes it possible to define the following parametrization for the solutions :

$$\varphi(\tau, u_r) = \varphi(x_\mu(\tau, u_r) - c_\mu(u_r)\tau.$$

so that all the interaction effects are included in c,(u). It implies, that the energy-momentum vector contains the interaction part of the Lagrangian and, hence, is a dynamical quantity. Since $c_\mu(u_r)$ is a velocity, it is directed along the surface normal. In this way, we shall seek a choice of $c_\mu(u_r)$ for which the Lorentz transformations are kinematical. Such a choice of c,(u) is equivalent to the following gauge fixing:

$$U_\mu = x_\mu(\tau, u_r) - \tau\eta_\mu(u) \approx 0.$$

Introducing then according to DIRAC [l0] new brackets for the arbitrary functions of dynamical variables one can convince oneself, that brackets for $x_\mu(\tau, u_r)$ and $\pi_\mu(\tau, u_r)$ become zero. These are therefore unphysical. The phase space is now defined by the conditions $U_\mu = V_\mu = 0$, which allows us to introduce in this gauge the Hamiltonian of the system.

### 3. Transformation of variables

We consider below the theory of the classical N-component field $\varphi_i(x)$, i == 1, ... , N, with action:
$$s = \int d^4x \mathcal{L}\left(\varphi_i(x), \partial_\mu\varphi_i(x)\right). \tag{1}$$
This action is a relativistically invariant functional. Noether's theorem implies the following conserved quantities (cf. also the Appendix for notational conventions):

$$P_\mu = \int d\sigma^\nu\, T_{\mu\nu} \tag{2}$$
$$M_{\alpha\beta} = \int d\sigma^\nu\, \xi_{\alpha\beta}{}^\mu T_{\mu\nu} - \left(I_{\alpha\beta}\right)_{ij}\varphi_j \frac{\partial \mathcal{L}}{\partial\partial^\nu\varphi_i(x)}, \tag{3}$$

in which

$$T_{\mu\nu} = \frac{\partial \mathcal{L}}{\partial \partial^\nu \varphi_i(x)} \frac{\partial \varphi_i}{\partial x^\mu} - g_{\mu\nu}\mathcal{L}$$

is the energy-momentum tensor. In relations (2) and (3) we have preserved the integration over the arbitrary surface to be free to choose a surface on which the Lorentz transformations possess kinematical character. We define the transformation of our field $\varphi_i(x)$ as follows:

$$\varphi_i(x) = S_{ik}(\vartheta(\tau))\tilde{\varphi}_\kappa(\bar{\Lambda}(\vartheta(\tau))(x(\tau,u) - q(\tau))) \tag{4}$$

and make a change of the integration variables in (1):
$$x_\mu = \Lambda_\mu{}^\nu(\bar{\Lambda}(\vartheta(\tau))\{\tau\alpha_\nu(u) - q_\nu(\tau)\}. \tag{5}$$
where the unit 4-vector $\alpha_\nu(u)$ is time-like. The field quantities $\tilde{\varphi}_\kappa(u,\tau)$ together with the scale deformation parameter $c(\tau)$ and the Poincare group parameters $\vartheta^{\alpha\beta}(\tau) = -\vartheta^{\beta\alpha}(\tau)$ and $q_\nu(\tau)$ provide the new set of dynamical variables. The surface metric tensor is given by the equation:
$$\gamma^{rs} = \frac{\partial \alpha_\nu(u)}{\partial u_r}\frac{\partial \alpha^\nu}{\partial u_s}.$$

A four-volume element in x-space is equal to the product of the 3-dimensional elementary volume in the $(u_1, u_2, u_3)$-space, given on the surface, and the elementary distance along the surface normal:

$$d^4x = \dot{c}(\tau)c(\tau)d^3u d\tau.$$
The action can then be written as follows:

$$S = \int d\tau L, \quad L = \int \dot{c}(\tau)c(\tau)d^3u d\tau \mathcal{L}\varphi_i(x), \partial_\mu\varphi_i(x)\big|_{\varphi=S\tilde{\varphi}} \tag{6}$$

where $L$ is the new Lagrangian of our system. To reexpress it in the new representation we need to calculate the derivatives $\partial_\mu \varphi_i(x)$, i. e.

$$\frac{\partial \varphi_i}{\partial x_\mu} = \left(\frac{\partial S_{ik}}{\partial \tau}\tilde{\varphi}_k + S_{ik}\frac{\partial \tilde{\varphi}_k}{\partial \tau}\right)\frac{\partial \tau}{\partial x_\mu} + S_{ik}\frac{\partial \tilde{\varphi}_k}{\partial u_r}\frac{\partial u_r}{\partial x_\mu}. \tag{7}$$

Differentiating equation (5), one obtains the following expressions for $\frac{\partial \tau}{\partial x_\mu}$ and $\frac{\partial u_r}{\partial x_\mu}$ :

Differentiating equation (5), one obtains the following expressions for $\frac{\partial \tau}{\partial x_\mu}$ and $\frac{\partial u_r}{\partial x_\mu}$ :

$$\frac{\partial \tau}{\partial x_\mu} = \frac{\alpha_\nu \bar{\Lambda}_\mu{}^\nu}{\alpha_\sigma \Omega^\sigma},$$
$$\frac{\partial u_r}{\partial x_\mu} = \bar{\Lambda}_\nu{}^\mu\left\{c(\tau)\frac{\partial \alpha^\nu(u)}{\partial u^r} + \frac{\alpha^\nu(u)}{\alpha_\sigma\Omega^\sigma}\frac{\partial \alpha^\rho(u)}{\partial u^r}\Omega^\rho\right\}$$

where the following notation is used for reasons of brevity:

$$\Omega_\sigma = \dot{c}(\tau)\alpha_\sigma + \bar{\Lambda}_\sigma{}^\nu \dot{q}(\tau) - \vartheta^{\alpha\beta}\bar{B}_{\alpha\beta}{}^{\gamma\delta}(\vartheta)\bar{\xi}_{\gamma\delta,\sigma}(\alpha).$$

The derivative $\dot{S}_{ik}$ is of the form:

$$\dot{S}_{ik} = S_{in}\dot{\vartheta}^{\alpha\beta}\bar{B}_{\alpha\beta}^{\gamma\delta}(\vartheta)(\bar{I}_{\gamma\delta})_{nk}.$$

We can now pass over to the Hamiltonian approach. Since the number of variables of the theory has been extended there are constraints. Indeed, by calculating the conjugate momenta, we obtain:

$$\tilde{\pi}(\tau,u) = \frac{\partial L}{\partial \dot{\tilde{\varphi}}(\tau,u)} = \dot{c}(\tau)c(\tau)S_{mn}\bar{\Lambda}_\nu^{\;\mu}\left.\frac{\partial \mathcal{L}}{\partial \partial^\mu \varphi}\right|_{\varphi=S\tilde{\varphi}}\alpha^\nu(u), \tag{8}$$

$$P_{\alpha\beta} = \frac{\partial L}{\partial \dot{\vartheta}^{\alpha\beta}(\tau,u)}\bar{B}_{\alpha\beta}^{\gamma\delta}(\vartheta)\int d^3u\left\{\bar{\xi}_{\gamma\delta}^{\;\sigma}(\alpha)\frac{\partial \alpha_\sigma}{\partial u_r}\frac{\partial \tilde{\varphi}_m}{\partial u^r}\tilde{\pi}_m - (\bar{I}_{\gamma\delta})_{mn}\tilde{\varphi}_n(\tau,u)\tilde{\pi}_m(\tau,u)\right\} \tag{9}$$

$$P_\mu = \frac{\partial L}{\partial \dot{q}^\mu(\tau)} = -\bar{\Lambda}_\nu^{\;\mu}\int d^3u\left\{\alpha^\nu(u)\mathcal{H} + c(\tau)\frac{\partial \alpha^\nu}{\partial u_r}\mathcal{P}_r\right\} \tag{10}$$

$$P_c = -\int d^3u \mathcal{H} \tag{11}$$

where

$$\mathcal{H} = \tilde{\pi}_m \tilde{\varphi}_{nl} - \dot{c}(\tau)c(\tau)\mathcal{L},$$

$$\mathcal{P}_r = \tilde{\pi}_m(\tau,u)\frac{\partial \tilde{\varphi}_m(\tau,u)}{\partial u^r},$$

$$\tilde{\varphi}_{nl} = \bar{\Lambda}_\nu^{\;\mu}S_{nm}\frac{\partial \varphi_m}{\partial x^\mu}\alpha^\nu. \tag{12}$$

It is now easy to find the following constraints:

$$\psi_\mu = P_\mu + \bar{\Lambda}_{\nu\mu}\int d^3u \alpha^\nu \mathcal{H} + c(\tau)\frac{\partial \alpha^\nu}{\partial u_r}\mathcal{P}_r, \tag{13}$$

$$\psi_{\alpha\beta} = \bar{L}_{\alpha\beta} - \int d^3u \Sigma_{\alpha\beta,m}(\tilde{\varphi}(\tau,u))\tilde{\pi}_m(\tau,u), \tag{14}$$

$$\psi_c = P_c + d^3u\mathcal{H}. \tag{15}$$

Here we used the notation:

$$\Sigma_{\alpha\beta,m}(\tilde{\varphi}(\tau,u)) = \bar{\xi}_{\gamma\delta}^{\;\sigma}(\alpha)c^2(\tau)\frac{\partial \alpha_\sigma}{\partial u_r}\frac{\partial \tilde{\varphi}_m}{\partial u^r} - (\bar{I}_{\gamma\delta})_{mn}\tilde{\varphi}_m(\tau,u).$$

One should note that in every specific model $\tilde{\varphi}_m$ has to be reexpressed in terms of the canonical variables.

## 4. The Conservation Laws

We should now express all physical quantities in the new representation and verify the validity of the conservation laws. We consider expressions (2) and (3). The surface element can be expressed in terms of the 4-volume element as follows:

$$d\sigma^\nu = \frac{\partial F(x)}{\partial x_\nu}\delta(F(x) - \tau_0)d^4x,$$

in which the function $F(x)$ determines the form of the surface in space-time. Substituting the transformations (4), (5) into the expressions of physical quantities and assuming

$$F(x) = x_\mu(\tau, u)\Lambda_\nu^{\ \mu}\alpha^\nu$$

we obtain after some manipulations:

$$P_\mu = \bar{\Lambda}_{\nu\mu}\int d^3u\, \alpha^\nu \mathcal{H} + c(\tau)\frac{\partial \alpha^\nu}{\partial u_r}\mathcal{P}_r, \tag{16}$$

$$M_{\alpha\beta} = -L_{\alpha\beta} - (J_{\alpha\beta})_{\mu\nu}q^\mu P_\nu. \tag{17}$$

It is known from the theory of Lie groups, that there exists a simple connection between the quantities $L_{\alpha\beta}$, and $\bar{L}_{\alpha\beta}$, indeed

$$L_{\alpha\beta} = \Lambda_\alpha^{\ \sigma}\Lambda_\beta^{\ \rho}\bar{L}_{\sigma\rho}. \tag{18}$$

Based on this relationship together with the constraint (14), we conclude, that the 4-angular-momenta is kinematical, whereas the 4-energy-momentum vector is a dynamical one. Taking the conservation laws into account one should note that because of the constraints the momenta $\tilde{\pi}_m(\tau, u)$ contain transverse and longitudinal parts concerning some direction defined by the solutions of the classical equations. The longitudinal part turns out to be unphysical and one can

$$\tilde{\varphi}_m(\tau, u) = g\tilde{\varphi}_{0m}(\tau, u) + \Phi_m(\tau, u), \quad \tilde{\pi}_m(\tau, u) = \tilde{\pi}_{mt}(\tau, u) + \tilde{\pi}_{ml}(\tau, u).$$

where $\tilde{\varphi}_{0m}(\tau, u)$ are the classical solutions, and $\Phi_m(\tau, u)$ together with the transverse momenta $\Phi_m(\tau, u)$ describe some fluctuations around them. The $\tilde{\pi}_{mt}(\tau, u)$ satisfy the following orthogonality conditions:

$$\int d^3u\, \Sigma_{\alpha\beta,m}(\tilde{\varphi}(\tau, u))\tilde{\pi}_{mt}(\tau, u) \approx 0. \tag{19}$$

They are, in fact, transversality conditions and will later be seen to play the role of constraints. For the longitudinal part $\tilde{\pi}_{ml}(\tau, u)$ we then obtain the following expression:

$$\tilde{\pi}_{nl}(\tau, u) = \frac{1}{g}N_n^{\alpha\beta}(\tau, u)\left(1 + \frac{1}{g}F\right)^{-1}_{\alpha\beta,\gamma\delta}\{\bar{L}^{\gamma\delta}(\vartheta(\tau)) - \int d^3u\, \Sigma^{\gamma\delta}_m(\tilde{\varphi}(\tau, u))\tilde{\pi}_{mt}(\tau, u)\}, \tag{20}$$

where the arbitrary functions $N^{\alpha\beta}_n(\tau, u)$ are normalized as follows:

$$\int d^3 u N^{\alpha\beta}_n(\tau, u) \Sigma^{\gamma\delta}_m (\varphi_{0m}(\tau, u)) = g^{\alpha\gamma} g^{\beta\delta}. \tag{21}$$

Besides,

$$F_{\alpha\beta,\gamma\delta} = \int d^3 u \Sigma_{\alpha\beta,n}(\Phi(\tau, u)\, N_{\gamma\delta}(\tau, u). \tag{22}$$

Thus the momenta $\tilde{\pi}_{mt}(\tau, u)$ and, consequently, the energy-momentum vector depends on the Lorentz group parameters through the quantities $\bar{L}_{\sigma\rho}$. At the same time the 4-angular-momenta $M_{\alpha\beta}$, coincide with the generators, but their Poisson brackets with the $\bar{L}_{\sigma\rho}$, is zero. This is just what we need because as a result of this and expression (15), we can develop the perturbation theory in inverse powers of $g$, which will be manifestly Lorentz-covariant in every order of the expansion.

## 5. Transition to the Quantum Theory

The considerations of the previous section allow us to pass to the Hamiltonian approach. It is easy to verify that the canonical Hamiltonian:

$$H = \dot{c}(\tau) P_c(\tau) + \dot{\vartheta}^{\alpha\beta}(\tau) P_{\alpha\beta}(\tau) + \dot{q}(\tau) P_c + \int d^3 u \dot{\varphi}_m(\tau, u) \tilde{\pi}_m(\tau, u) - L$$

is identically zero upon insertion of the constraints. Following Dirac's theory, we take the Hamiltonian as a linear combination of constraints. i.e.

$$H' = a(\tau) \Psi_c + a_\mu(\tau) \Psi^\mu + b_{\alpha\beta} \Psi^{\alpha\beta}. \tag{23}$$

The equation of motion for any particular dynamical variable has the form

$$f(\dot{\tau}) = \frac{\partial f}{\partial t} + \{f, H'\} \tag{24}$$

and lead for the constraints (they must be time independent) to the following consistency conditions;

$$a(\tau)\{\Psi_c, \Psi_c\} + a_\mu(\tau)\{\Psi_c, \Psi^\mu\} + b_{\alpha\beta}(\tau)\{\Psi_c, \Psi^{\alpha\beta}\} \approx 0,$$

$$a(\tau)\{\Psi^\nu, \Psi_c + a_\mu(\tau)\{\Psi^\nu, \Psi^\mu\} + b_{\alpha\beta}(\tau)\{\Psi^\nu, \Psi^{\alpha\beta}\} \approx 0,$$

$$a(\tau)\{\Psi^{\rho\sigma}, \Psi_c\} + a_\mu(\tau)\{\Psi^{\rho\sigma}, \Psi^\mu\} + b_{\alpha\beta}(\tau)\{\Psi^{\rho\sigma}, \Psi^{\alpha\beta}\} \approx 0.$$

One can verify, that the constraints form a closed algebra with constant structure coefficients so that the consistency conditions become the identities: $0 = 0$. This means that our theory does not contain constraints except (13)-(15). At the same time, there are the arbitrary coefficients $a(\tau)$, $a_\mu(\tau)$ and $b_{\alpha\beta}(\tau)$ in the theory, which can be connected with the velocities of the parameters, introduced by transformations (4) and (5). Indeed from the equation of motion, we obtain:

$$a(\tau) = \dot{c}(\tau), \quad a_\mu(\tau) = \dot{q}_\mu(\tau), \quad b_{\alpha\beta}(\tau) = \bar{B}^{\gamma\delta}_{\alpha\beta}(\vartheta) \dot{\vartheta}_{\gamma\delta}(\tau).$$

The arbitrariness can be avoided by introducing the gauge conditions. Before we fix these conditions we rewrite the constraints:

$$\Psi_c = P_c + \int d^3 u \mathcal{H}, \tag{25}$$

$$\Psi_\mu = \Lambda_\mu^\nu P_\nu + \int d^3 u \left( \alpha_\mu \mathcal{H} + c(\tau) \frac{\partial \alpha_\mu}{\partial u_r} \mathcal{P}_r \right), \tag{26}$$

$$\Psi_{\alpha\beta} = \int d^3 u \Sigma^{\gamma\delta}{}_m \left( \varphi_{0m}(\tau, u) \right) \pi_{nt}(\tau, u) \tag{27}$$

We recall that the variables

$$c(\tau), \ P_c(\tau), \ q_\mu(\tau), \ P_\mu(\tau), \ \vartheta_{\alpha\beta}(\tau), \ P_{\alpha\beta}(\tau), \Phi_k(\tau, u), \pi_{nt}(\tau, u) \tag{28}$$

constitute the phase space of theory. We next introduce the following gauge conditions:

$$\chi_c = c(\tau) - \tau, \tag{29}$$

$$\chi_\mu = \Lambda_\mu^\nu q_\nu(\tau) - \tau P_\mu(\tau), \tag{30}$$

$$\chi_{\alpha\beta} = \int d^3 u N_{\alpha\beta,n}(\tau, u) \Phi_n(\tau, u). \tag{31}$$

These conditions are chosen in such a way, that their Poisson brackets with each other become zero (such a choice is convenient, but not necessary). The second requirement we impose on the gauge conditions is that the matrix composed of Poisson brackets of constraints with gauge conditions should be nonsingular. We now modify the Hamiltonian dynamics with the help of the brackets introduced by DIRAC [l0]. We introduce the notation $A_{ab} = \{\Psi_a, \chi_b\}$, where $\Psi_a$ and $\chi_b$ are the sets of constraints and gauge conditions respectively. The new brackets for any dynamical quantity are now defined as follows:

$$\{f, h\}_D = \{f, h\} - \{f, \Psi_a\} A_{ba}^{-1} \{\chi_{b,h}\} - \{f, \chi_a\} A_{ab}^{-1} \{\Psi_b, h\}$$

where the nonsingular matrix A can be represented in the form

$$A = \begin{pmatrix} g & \{\Psi, \chi_c\} & \{\Psi, \chi_{\alpha\beta}\} \\ 0 & -1 & \{\Psi_c, \chi_{\alpha\beta}\} \\ 0 & 0 & -I \end{pmatrix}$$

where g is the metric tensor of 4-space-time, $\Psi = \Psi_\mu$ , and $I$ is a unit 6 x6 matrix. The inverse matrix $A^{-1}$ has the following form:

$$A^{-1} = \begin{pmatrix} -g & -\{\Psi, \chi_c\} & \{\Psi, \chi_{\alpha\beta}\} + \{\Psi, \chi_c\} \cdot \{\Psi_c, \chi_{\alpha\beta}\} \\ 0 & -1 & g\{\Psi_c, \chi_{\alpha\beta}\} \\ 0 & 0 & -I \end{pmatrix}.$$

Evaluating the Dirac brackets for our canonical variables, we find that $c(\tau), P_c(\tau), q_\mu(\tau), P_\mu(\tau)$ become unphysical; indeed:

$$\{c(\tau), P_c(\tau)\}_D = \{q_\mu(\tau), P_\nu(\tau)_D = 0,$$

whereas the Dirac brackets for the other canonical pairs have the following form:

$$\{\vartheta_{\alpha\beta}(\tau), P_{\gamma\delta}(\tau)\}_D = g_{\alpha\gamma} g_{\beta\delta}, \ \alpha > \gamma, \beta > \delta, \tag{32}$$

$$\{\Phi_m(\tau, u), \pi_{nt}(\tau, u')\}_D = \delta_{mn}\delta^3(u - u') - \Sigma_{\alpha\beta,n}(\varphi_0(\tau, u)) N^{\alpha\beta}{}_m(\tau, u') \equiv A_{mn}(u, u'). \tag{33}$$

We have to choose a quantity, which generates the evolution of the system. One can verify, that the brackets of dynamical variables (or physical quantities) with $P_c$ provide the evolution in $\tau$. For example

$$\{P_c, \Phi_m(\tau, u)\}_D = \int d^3 v A_{mn}(u, v) \frac{\delta}{\delta \pi_{nt}(\tau, v)} \int d^3 u' \mathcal{H}, \tag{34}$$

$$\{P_c, \pi_{nt}(\tau, u)\}_D = \frac{\delta}{\delta \pi_{nt}(\tau, v)} \int d^3 u' \mathcal{H}. \tag{35}$$

Due to constraints we can eliminate unphysical degrees of freedom and define the Hamiltonian - which generates the evolution of the system - as follows:

$$H = \int d^3 u \mathcal{H}.$$

In general, we have to add to this Hamiltonian the linear combination of remaining primary constraints $\Psi_{\alpha\beta}$, but their Dirac brackets with dynamical quantities become zero, so we can get these constraints in a strong sense. Therefore the equations of motion have the form:

$$\dot{f} = \frac{\partial f}{\partial \tau} + \{f, H\}_D. \tag{36}$$

The transition to quantum theory is achieved with the help of the new brackets. In particular, the commutator of two operators $f, g$ is defined by the relation:

$$[f, g] = i\{f, H\}_D. \tag{37}$$

From conditions (32) and (33) we can obtain the algebra of the operators, realized in the Hilbert space, in which the variables $P_{\alpha\beta}$ and $\pi_{nt}(\tau, v)$ are acting as the following differentiation operators:

$$P_{\alpha\beta} = -i\frac{\partial}{\partial \vartheta^{\alpha\beta}}, \qquad \pi_{mt}(\tau, v) = -i\int d^3 v A_{mn}(u, v) \frac{\delta}{\delta \Phi_{nt}(\tau, v)}, \tag{38}$$

he space is a direct sum of two spaces, realizing the representation (38).

Returning to expression (20), we can conclude that the state vector is an eigenvector of the two commuting operators, $H$ and $\bar{L}_{\alpha\beta}$ because $[H, \bar{L}_{\alpha\beta}] = 0$. Hence we can separate the $\vartheta^{\alpha\beta}$-dependence of the state vector:

$$\psi(\vartheta, \Phi(\tau, u) = D(\vartheta)\psi'(\Phi(\tau, u)). \tag{39}$$

Hence, the quantum equations for the remaining fluctuation operators i.e.

$$i\frac{\partial \Phi_m(\tau, u)}{\partial \tau} = [\Phi_m(\tau, u), H], \quad i\frac{\partial \pi_{mt}(\tau, v)}{\partial \tau} = -[\pi_{mt}(\tau, v), H]. \tag{40}$$

are explicitly covariant. Expanding the momenta $\pi_{mt}(\tau, v)$ (and correspondingly $H$)) in a series of inverse powers of the coupling constant $g$, we can construct the exact covariant perturbation theory for the equations (40) to every order of the expansion. Before expanding it is necessary to transform the state vector as follows:

$$\psi' \Rightarrow \exp\left(ig\int d^3u s_m(\tau, u)\Phi_m\right)\psi' \tag{41}$$

provided that

$$\int d^3u \Sigma_{\alpha\beta,m}(\varphi_0(\tau, u))s_m(\tau, u) = 0. \tag{42}$$

If the $s_m(\tau, u)$ do not satisfy (42), we can define another $s_m'(\tau, u)$ which would satisfy this requirement, indeed

$$s_m'^{(\tau,u)} = s_m(\tau, u) - N_{\alpha\beta,m}(\tau, u)\int d^3v \Sigma_{\alpha\beta,m}(\varphi_0(\tau, v))s_m(\tau, v).$$

The transformation (41) implies, that the quantum operator $\pi_{mt}(\tau, v)$ has to be replaced by

$$\pi_{mt}(\tau, v) \Rightarrow gs_m(\tau, u) + \pi_{mt}(\tau, v).$$

We then expand the momenta $\pi_{mt}(\tau, v)$ and the Hamiltonian $H$ in a series of inverse powers of $g$, i.e.

$$\pi_{mt}(\tau, v) = g\pi_{0mt}(\tau, v) + \pi_{1mt}(\tau, v) + g^{-1}\pi_{2mt}(\tau, v) + \cdots \tag{43}$$

$$H = g^2 H_0 + gH_1 + H_2 + g^{-1}H_3 + \cdots \tag{44}$$

The dominant terms of these series are purely classical. The terms that follow are linear ($\sim g$) and quadratic ($\sim g^0$) in the fluctuation variables. Thus, the quantum equations up to the one-loop approximation are:

$$i\frac{\partial \varphi_{0n}(\tau, u)}{\partial \tau} = [\Phi_n(\tau, u), H_1],$$

$$i\frac{\partial \pi_{0nt}(\tau, u)}{\partial \tau} = [\pi_{1nt}(\tau, u), H_1],$$

$$i\frac{\partial \pi_{0nt}(\tau, u)}{\partial \tau} = [\pi_{1nt}(\tau, u), H_1],$$

$$i\frac{\partial \Phi_n(\tau, u)}{\partial \tau} = [\Phi_n(\tau, u), H_1],$$

$$i\frac{\partial \pi_{1nt}(\tau, u)}{\partial \tau} = [\pi_{1nt}(\tau, u), H_2] + [\pi_{2nt}(\tau, u), H_1].$$

We see, that the first line reproduces the classical equations of motion whereas the others coincide with one-loop approximation equations. It should be noted that one has to be careful in manipulating quantum variables, but up to the one-loop approximations no operator ordering problem arises, and hence this question does not concern us here.

**Appendix**

In this appendix we summarize some properties of the Lorentz group which we use in the text. The Lorentz transformation matrix satisfies the following orthogonality and product conditions :

$$\Lambda_\alpha^\beta \bar{\Lambda}_\beta^\gamma = \delta_\alpha^\gamma, \qquad \Lambda(\vartheta_1)\Lambda(\vartheta_2) = \Lambda(\vartheta_3). \tag{A.1}$$

From the latter condition, we deduce that

$$\vartheta_2^{\gamma\delta} = \varphi^{\gamma\delta}(\theta_1, \theta_2).$$

The matrices satisfy the equations:

$$\frac{\partial \Lambda(\vartheta)}{\partial \vartheta^{\alpha\beta}} = B_{\alpha\beta}^{\gamma\delta}(\theta) J_{\gamma\delta} \Lambda(\vartheta),$$

$$\frac{\partial \bar{\Lambda}(\vartheta)}{\partial \vartheta^{\alpha\beta}} = \bar{B}_{\alpha\beta}^{\gamma\delta}(\theta) \bar{J}_{\gamma\delta} \Lambda(\vartheta). \tag{A.2}$$

Where

$$B_{\alpha\beta}^{\gamma\delta}(\vartheta_2) = \left.\frac{\partial \varphi^{\gamma\delta}(\theta_1, \theta_2)}{\partial \vartheta_1^{\alpha\beta}}\right|_{\vartheta_1 = \vartheta_2^{-1}},$$

$$\bar{B}_{\alpha\beta}^{\gamma\delta}(\vartheta_1) = \left.\frac{\partial \varphi^{\gamma\delta}(\theta_1, \theta_2)}{\partial \vartheta_2^{\alpha\beta}}\right|_{\vartheta_2 = \vartheta_1^{-1}} \tag{A.3}$$

which become unit matrixes for zero arguments. Then

$$J_{\alpha\beta} = \left.\frac{\partial \Lambda(\vartheta)}{\partial \vartheta^{\alpha\beta}}\right|_{\vartheta=0},$$

$$\bar{J}_{\alpha\beta} = \left.\frac{\partial \bar{\Lambda}(\vartheta)}{\partial \vartheta^{\alpha\beta}}\right|_{\vartheta=0} \tag{A.4}$$

We define matrices $A, \bar{A}$ which are inverse to $B, \bar{B}$ respectively by equations:

$$AB = I, \quad \bar{A}\bar{B} = I, \tag{A.5}$$

I am a unit matrix. With the help of these matrices, we can construct the following quantities:

$$L_{\alpha\beta} = B_{\alpha\beta}^{\gamma\delta} P_{\gamma\delta}, \quad \bar{L}_{\alpha\beta} = \bar{B}_{\alpha\beta}^{\gamma\delta} P_{\gamma\delta}. \tag{A.6}$$

They obey the usual Poisson brackets for the Lorentz group parameters¨

$$\{L_{\alpha\beta}, L_{\gamma\delta}\} = C_{\alpha\beta,\gamma\delta}^{\sigma\rho} L_{\sigma\rho}, \quad \{\bar{L}_{\alpha\beta}, \bar{L}_{\gamma\delta}\} = -C_{\alpha\beta,\gamma\delta}^{\sigma\rho} \bar{L}_{\sigma\rho}, \quad \{L_{\alpha\beta}, \bar{L}_{\gamma\delta}\} = 0. \tag{A.7}$$

where $C_{\alpha\beta,\gamma\delta}^{\sigma\rho}$ are well-known structure constants.

We are interested in the Minkowsky space coordinate transformations :

$$x_\mu \Rightarrow x'_\mu = \Lambda_\mu^{\ \nu} x_\nu \equiv f_\mu(x, \vartheta). \tag{A.8}$$

From the product law we can obtain the following equations:

$$\frac{\partial x'_\mu}{\partial \vartheta^{\alpha\beta}} = B_{\alpha\beta}^{\gamma\delta}(\vartheta) \xi_{\alpha\beta,\mu}(x') \tag{A.9}$$

with

$$\xi_{\alpha\beta,\mu}(x') = \left.\frac{\partial f_\mu(x,\vartheta)}{\partial \vartheta^{\alpha\beta}}\right|_{\vartheta=0}. \tag{A.10}$$

Inverting the transformations we obtain the equations¨

$$\frac{\partial x_\mu}{\partial \vartheta^{\alpha\beta}} = \bar{B}_{\alpha\beta}^{\gamma\delta}(\vartheta) \bar{\xi}_{\alpha\beta,\mu}(x) \tag{A.9}$$

with

$$\bar{\xi}_{\alpha\beta,\mu}(x) = \left.\frac{\partial \bar{f}_\mu(x,\vartheta)}{\partial \vartheta^{\alpha\beta}}\right|_{\vartheta=0}. \tag{A.10}$$

The quantities

$$X_{\alpha\beta} = \xi_{\alpha\beta,\mu}(x)P^{\alpha\beta}, \quad \bar{X}_{\alpha\beta} = \bar{\xi}_{\alpha\beta,\mu}(x)P^{\alpha\beta} \tag{A.11}$$

with $\{x_\mu, P_\sigma\} = g_{\mu\sigma}$ satisfy the conditions similar to (A.7). These formulae give two different representations of the homogeneous Lorentz group. The matrix $S_{ik}(\vartheta)$, determines the transformation properties of the field variables $\varphi_\iota(x)$. They obey the equation

$$\frac{\partial S_{ik}(\vartheta)}{\partial \vartheta^{\alpha\beta}} = B_{\alpha\beta}^{\gamma\delta}(\theta)(I_{\gamma\delta})_{in} S_{nk}(\vartheta), \quad I_{\gamma\delta} = \left.\frac{\partial S(\vartheta)}{\partial \vartheta^{\gamma\delta}}\right|_{\vartheta=0} \tag{A.12}$$

If $\bar{S}$ are inverse to $S$, then

$$\frac{\partial \bar{S}_{ik}(\vartheta)}{\partial \vartheta^{\alpha\beta}} = \bar{B}_{\alpha\beta}^{\gamma\delta}(\theta)(\bar{I}_{\gamma\delta})_{in} S_{nk}(\vartheta), \quad \bar{I}_{\gamma\delta} = \left.\frac{\partial \bar{S}(\vartheta)}{\partial \vartheta^{\gamma\delta}}\right|_{\vartheta=0} . \tag{A.13}$$